\pdfoutput=1
\documentclass[b5paper, 12pt, twoside, openright]{AAN}  
\graphicspath{{Figures/}}  
\usepackage[square, numbers, comma, sort&compress]{natbib} 
\usepackage{verbatim} 
\usepackage{vector}  
\usepackage{amsmath}
\usepackage{amsfonts}
\usepackage{amssymb}
\usepackage{graphicx} 
\usepackage{polski}
\usepackage[utf8]{inputenc}
\usepackage{indentfirst} 
\usepackage{enumitem}
\usepackage{textcomp}
\usepackage{wasysym}
\usepackage{url}
\usepackage{marvosym} 
\usepackage{lipsum}
\begin{document}

\setstretch{1}  
\mainmatter	  
\chapter*{\vspace{-2cm}{\LARGE R Scuti: Close Alternating Pulsation Periods or Chaos in the RV Tau -- Type Star?}}
\addtocontents{toc}{\vspace{-1em}}
\addcontentsline{toc}{chapter}{{\normalsize Author -- {\normalfont {\it Title}}}} 
\fancyhf{}
\fancyhead[LE]{L.S. Kudashkina, I.L.Andronov}
\fancyhead[RE]{\thepage}
\fancyhead[RO]{R Scuti}
\fancyhead[LO]{\thepage}
\thispagestyle{myheadings}

\hspace{-0.7cm}{\Large \textbf{L.S. Kudashkina, I.L.Andronov}}\\
\\
{\normalsize 
\hspace{-0.7cm}Department "Mathematics, Physics and Astronomy"\\ 
Odessa National Maritime University, Odessa 65029, Ukraine
}
\\

\hspace{-0.7cm}{\bf Abstract}\\
Results of analysis of 60\,010 data photometric observations from the AAVSO international database are presented, which span 120 years of monitoring. The periodogram analysis shows the best fit period of 70.74d, a half of typically published periods for smaller intervals. Contrary to expectation for deep/shallow minima, the changes between them are not so regular. There may be series of deep (or shallow) minima without alternations. There may be two acting periods of 138.5 days and 70.74, so the beat modulation may be expected. The dependence of the phases of deep minima argue for two alternating periods with a characteristic life-time of a mode of 30years. These phenomenological results better explain the variability than the model of chaos.

\section*{Introduction}
R Sct = BD-05 4760= IRAS 18448-0545= HIP 92202. This star belongs to RV Tauri - type variables (RVA). RV Tauri stars are generally considered to be post-AGB stars with low initial masses (Jura, 1986). They are metal-poor supergiants of intermediate spectral type which show a pulsation variability with a light curve characterized by alternating deep and shallow minima (Mantegazza, 1991). The abundance ratios show that they have experienced first dredge-up at the bottom of the red giant branch. Based on their infrared dust excesses, RV Tau stars are classified into two groups: those with extensive warm dust and those without evidence of dust in the near-infrared region. R Sct is the brightest star in the visible in the latter group. R Sct has a reported period of 147 days. The effective temperature varies from 4750 to 5250 K; the
spectral type may vary as late as M3 at minimum phase (Matsuura et al., 2002).

First paper on the star, which is listed in the ADS, is dated 1890, and is devoted to the description of the spectrum (Espin, 1890). The GCVS (Samus et al., 2017) provides ephemeris with different periods for 15 light curve intervals. The smallest value of the period that occurs is $137.4^{\rm d}$, and the largest is $152.5^{\rm d}$.

\section*{Observations}
For the analysis, we have used the data published by the American Association of Variable Stars observers (AAVSO, Kafka 2020). We have used long-term photometrical observations - visual ones and that with the filter V. After the cleaning the data by removing outliers, 60\,010 data points remained in the range JD\,2414459.4 -- 2458470.851 (1898--2018), totally, 120 years of observation. As the star shows complicated photometric behaviour typical to the RV Tau-type, it got this classification (Samus' et al., 2017).

Earlier, we have already investigated photometric behavior of several RV Tauri stars. In total, more than 40 objects of the RVA and RVB types were studied, their periodogram analysis was carried out (Kudashkina et al., 1998; Kudashkina, 2019; Kudashkina, 2020a), the mean light curves (Kudashkina,2020b; Kudashkina, 2020c) and phase portraits (Kudashkina \& Andronov, 2017a; Kudashkina \& Andronov, 2017b) were plotted, the behavior of the mean brightness, amplitude and phase with time (Kudashkina et al., 2013). For almost all the stars studied, fairly regular mean curves were obtained over the entire observation interval, which were then smoothed by a trigonometric polynomial, and thus an atlas of mean curves was created (Andronov \& Chinarova, 2003), similar to the atlas of mean curves for Mira Ceti-type stars (Kudashkina \& Andronov, 1996). But not in the case of R Scuti! For this object, the maximum peak on the periodogram does not reflect the star's variability over the entire observation interval, and the average light curve with this period is very "smeared".

\section*{Periodogram Analysis}
Periodogram Analysis was made using the trigonomeric polynomial least square approximations of orders $s=1,$ 2 and 4. The test function $S(f)$ is the ratio of the variance of the approximation to the variance of observations (Andronov 1994, 2003, 2020; Andronov et al., 2020). The algorithm was realized in the software MCV (Andronov and Baklanov, 2004).  
\begin{equation}
    x(t)=C_1+\sum_{j=1}^s (C_{2j}\cos(j\omega t)+C_{2j+1}\sin(j\omega t)).
\end{equation}
Few values of $s$ are used to test for possible multi-harmonic periodic variations.

\begin{figure}
    \centering
    \includegraphics[width=6.1in,height=3.95in]{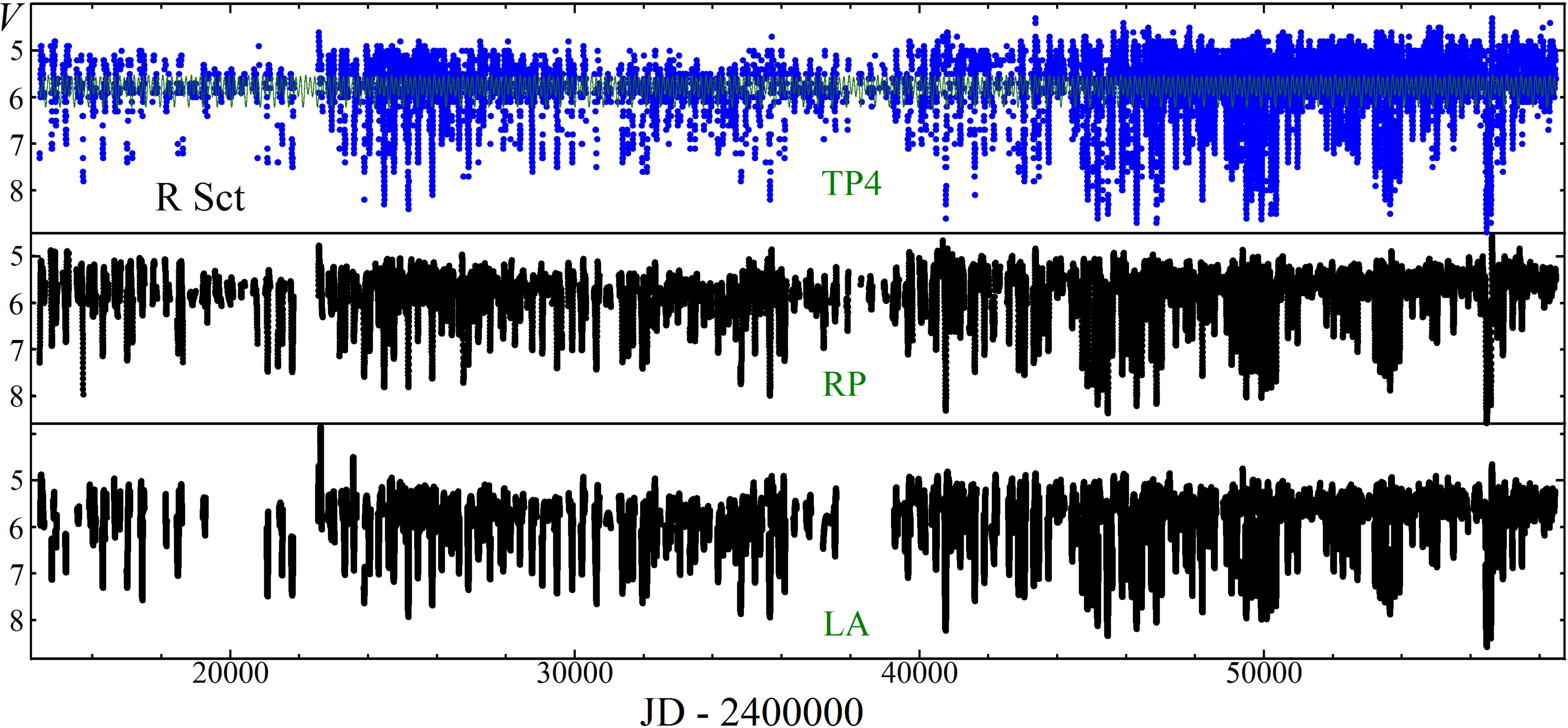}
    \caption[Fig.]{Observations of R Sct from the AAVSO database (blue circles) and its approximations using the trigonometric polynomial of order $s=4$ (TP4, up), running parabola (RP, middle), local approximations (LA, bottom).}
    \label{fig:1}
\end{figure}

\begin{figure}
    \centering
    \includegraphics[width=6.1in,height=3.95in]{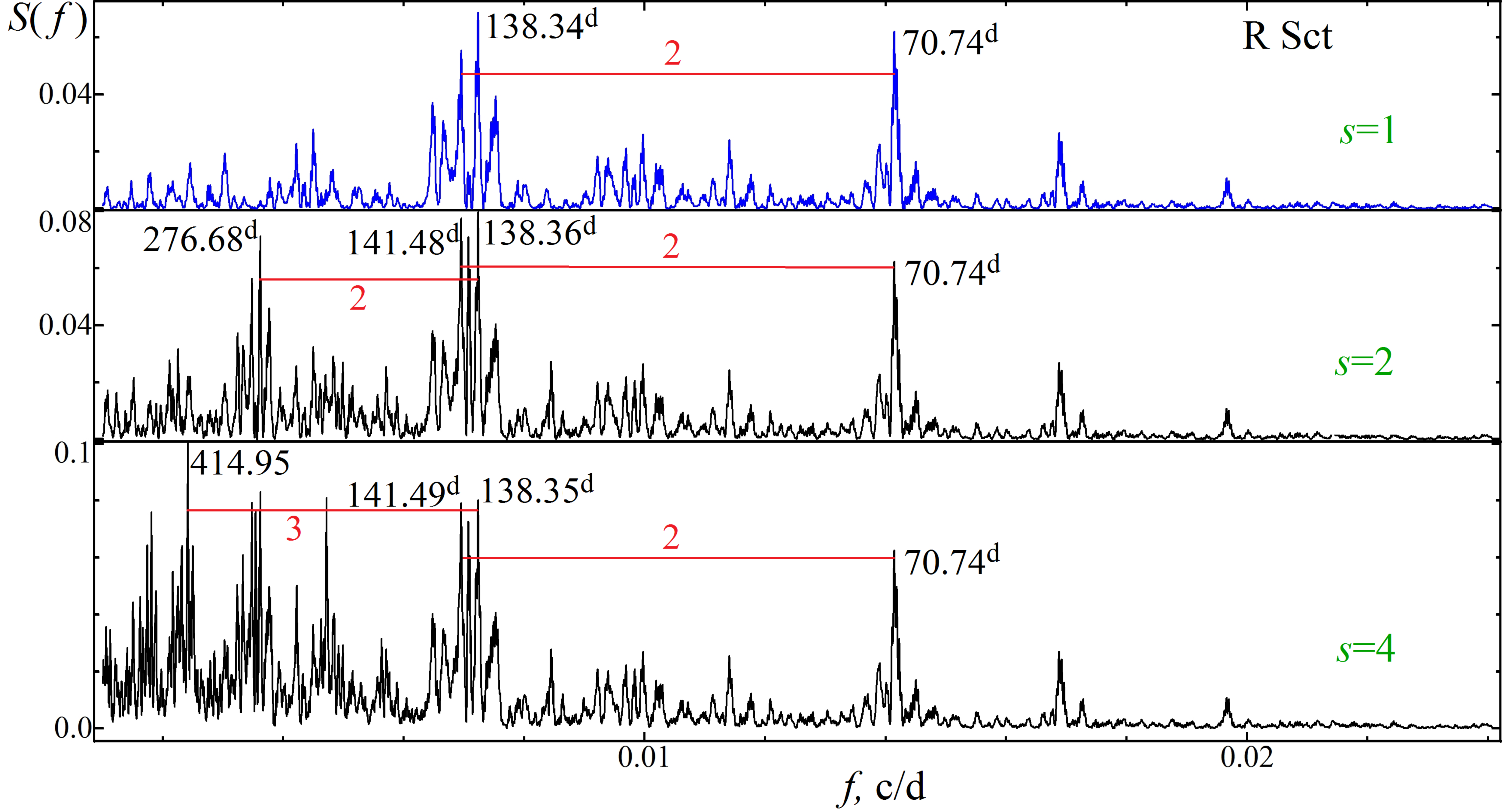}
    \caption[Fig.]{Periodogram $S(f)$ of R Sct obtained using the trigonometric polynomial of orders $s=1,$  2, 4. The highest peaks are marked with corresponding periods. The horizontal red lines show pairs of peaks with an integer ratio 2 or 3.}
    \label{fig:2}
\end{figure}

The results of the periodogram analysis are shown in Fig. 2. The best period for the entire data set is $138.34^{\rm d}$ for $s=1.$
The peak for $s=4$ corresponds to the elements
\begin{equation}
Min.JD=2447423.07+414.95\cdot E
\end{equation}
The corresponding phase light curve is shown in Fig. 3. It shows drastic phase shifts, whereas the approximation corresponds to a stable light curve. As there was a peak close to $70^{\rm d},$ which is 6 times smaller than the apparent value $414.95^{\rm d}$, we also made the approximation with $s=6.$ The corrected period is
$P=415.03\pm.0.01^{\rm d},$ and the mean
minimum is shifted by $0.02P.$
Thus both curves are shown in Fig. 3.

However, it is not possible to obtain a good mean light curve with this period. In addition to the $70.74^{\rm d}$ period, which is close to the 2: 1 (1.96) ratio, there is also a peak at the $276.68$ period value, which is a doubling of the formal period and is 3.91 with a $70.74^{\rm d}$ period. We have not found anywhere in the literature mentions of a period twice as long as the formal one. However, with this longest period, the best mean light curve is obtained. It is clearly seen how the phase of deep brightness minima changes (Fig. 3).

\section*{Scalegram Analysis}

Scalegram analysis was made using the algorithm of the weighted running parabola approximation (RP) introduced by Andronov (1987) and extended by Andronov (2003). The corresponding test-functions are shown in Fig. 4.
The maximum of the S/N=SNR (signal-to-noise ratio) occurs at $\Delta t=32^{\rm d}.$ The corresponding approximation is shown in Fig. 1. Contrary to the multi-harmonic  approximation with a constant period and constant shape of the phase curve, the RP approximation follows significant changes of the amplitude. The analysis of the light curve shows deep and shallow minima, as well as humps.

The $\Lambda-$ scalegram Andronov (2003) shows three peaks corresponding to ''periods'' of $P_{\Lambda 1}=70^{\rm d}$ (and a corresponding effective semi-amplitude $R_{\Lambda 1}=446$ mmag), $P_\Lambda=135^{\rm d}$ ($R_\Lambda=460$ mmag). Much weaker peak corresponds to $P_{\Lambda 3}=284^{\rm d},$ $R_{\Lambda 3}=274$ mmag. 

\begin{figure}
    \centering
    \includegraphics[width=6.1in]{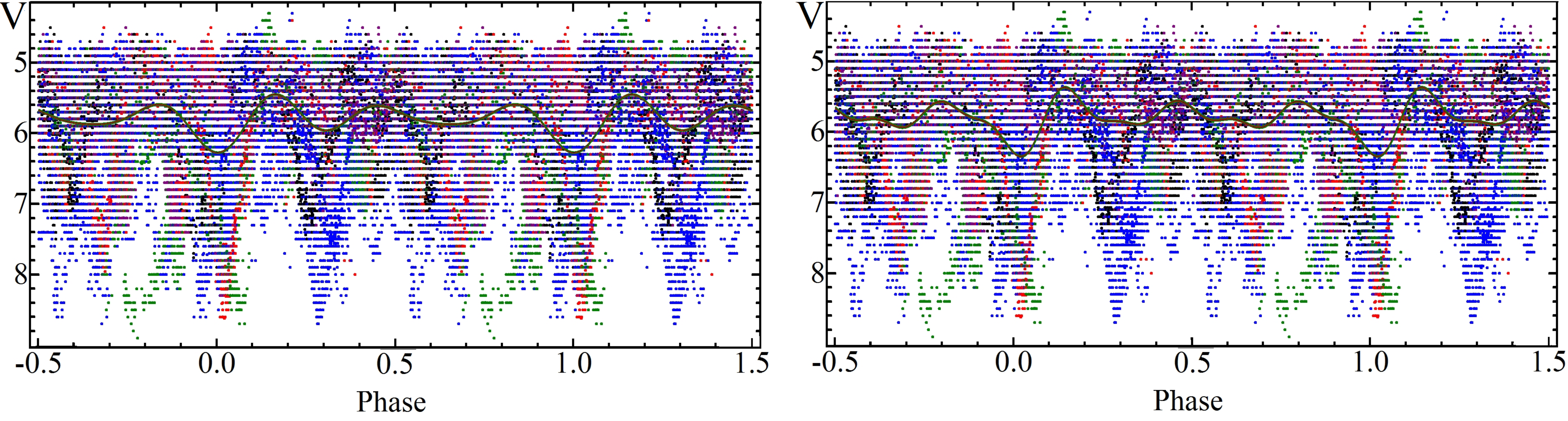}
    \caption[Fig.]{Phase light curves for the trigonometric polynomial approximations of degree $s=4$ (left) and $s=6$ (right). Observations are shown as circles, the approximations - by lines,}
    \label{fig:3}
\end{figure}

\begin{figure}
    \centering
    \includegraphics[width=6.1in]{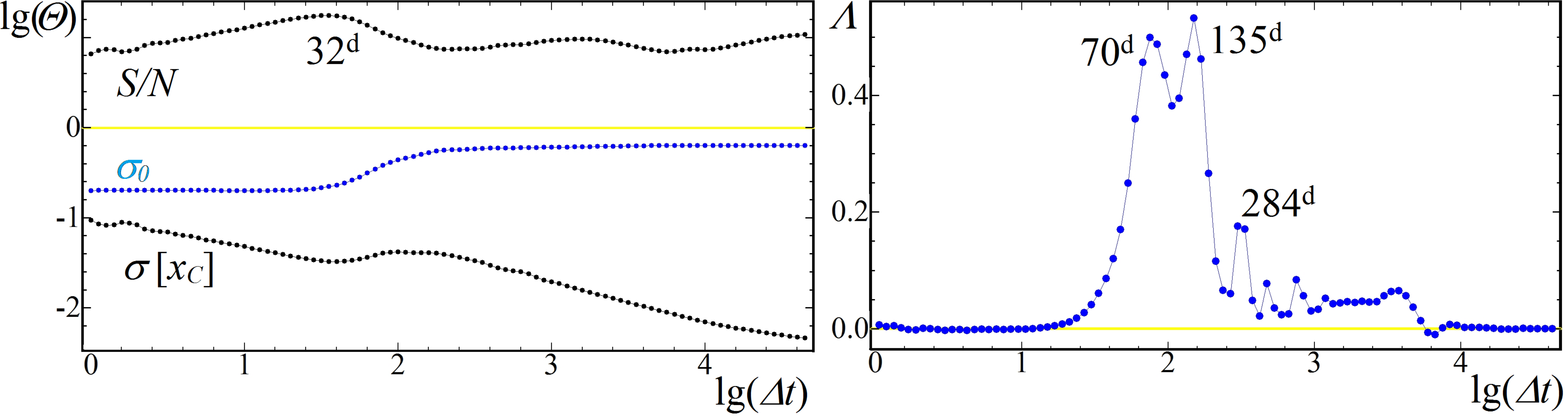}
    \caption[Fig.]{Left: The scalegrams $\sigma_0,$ $\sigma[x_C],$ S/N. Right: $\Lambda(\Delta t)$ scalegram. The numbers correspond to the filter half-width $\Delta t$ (left) and ''effective periods'' (right).}
    \label{fig:4}
\end{figure}

\section*{Times of Minima}

Looking for complicated behaviour of the light curve, we have made determination of individual extrema. 
For this, we have used the software MAVKA (Andrych 2020, 
The near-extremal intervals were marked, with a total number 609.

These data were approximated with a polynomial of statistically optimal degree, similar to the compilation of the catalogue of the individual characteristics of pulsations of semi-regular stars (Chinarova \& Andronov, 2000). Other methods like ''the asymptotic parabola'' (Marsakova and Andronov 1996, Andronov, 2005; Andrych et al. 2015) and ''parabolic spline'' (Andrych et al. 2020a). (Andrych et al. 2020b) tested effectivity of different methods. As the observations are distributed very irregularly, many of the extrema have short intervals. So finally we decided to use a single method - polynomials, which are characterized by smaller number of parameters.

The dependence of the height of the maxima and the minima on time is shown in Fig. 5. It shows relatively smaller scatter for the maxima from $3.85^{\rm m}$ to $6.46^{\rm m}$ and a larger one from $5.10^{\rm m}$ to $8.58^{\rm m}$ for the minima. 

These ranges partially overlap, so the data are shown at different panels. Moreover, one may suggest a binary character of the brightness of minima. So we classify the minima as faint $(m\ge 6.45^{\rm m})$ and bright ones.   

The time intervals between the maxima $\delta t_{\rm max}$ (separately, also minima $\delta t_{\rm min}$) are shown in Fig. 6. The range of values is $18^{\rm d}-2232^{\rm d}$ and $16^{\rm d}-1813^{\rm d}$ for the maxima and minima, respectively. The upper limit is due to the gap of observations more than a century ago, during and near the World War I, and thus has no physical meaning.

The intervals show condensation to the values $P\sim70^{\rm d}$ and  its integer multipliers. However, there are many points between the expected equidistant horizontal lines. 
The ''multiple'' periods at this diagram are due to missing extrema - some because they are within error corridor of the observations, some because of missing observations due to invisibility of the star at the day.
The scatter is rather large, showing phase drifts of the individual cycles.

As the observations show such strong period/phase changes, we apply another method. 

At first, we have decreased the number of minima timings to 129, using only ''deep'' minima with a brightness $m\ge 6.45^{\rm m}.$ The ''shallow'' minima may be partially hidden/biased by observational errors.
Additionally, we have used seven moments of minima  published by H\"ubscher (2011). They are close to our results, and unfortunately fill no gaps in the AAVSO data. Similarly, we tried to fill the gap with ''shallow'' minima near JD 2437000. They show larger scatter, and do not improve cycle number count.

\begin{figure}[bt]
    \centering
    \includegraphics[width=6.1in]{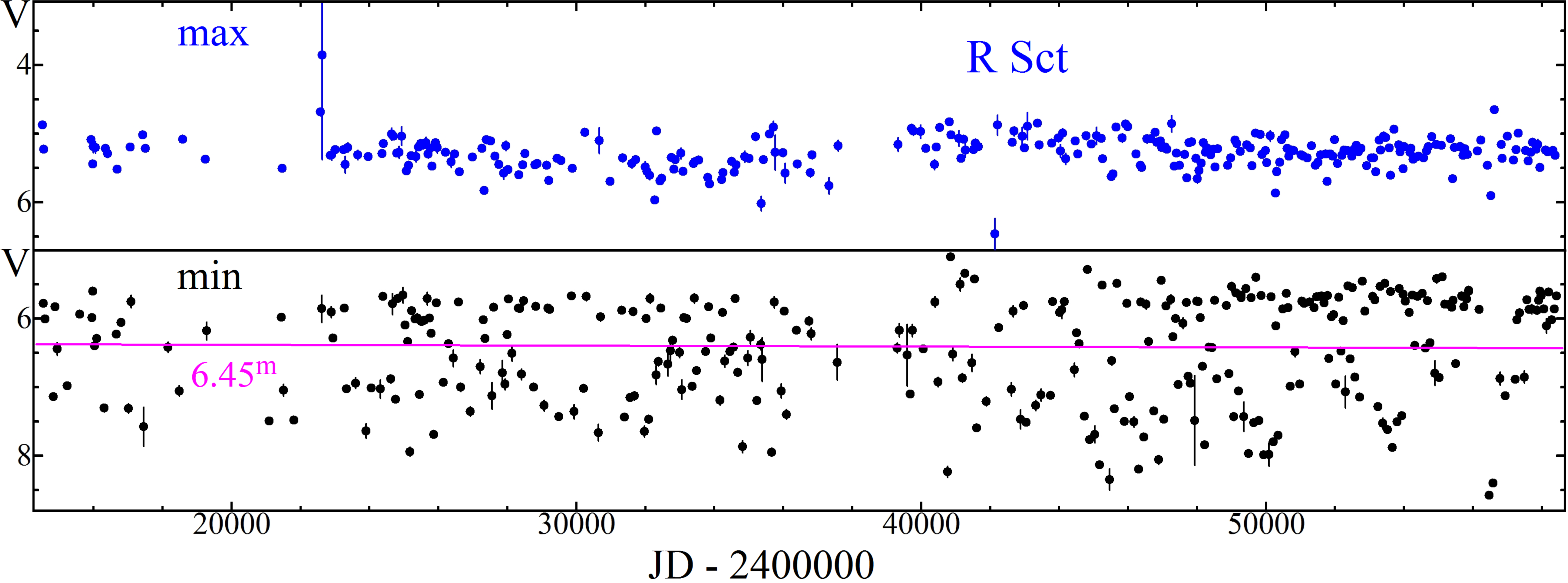}
    \caption[Fig.]{The brightness at maxima and minima vs. time. While the maxima may show a smoth wave with $\sim 20,000^{\rm d}$, the minima show different types (''shallow and deep'', or ''bright and faint''). The violet line shows a border between these two types.}
    \label{fig:5}
\end{figure}

\begin{figure}
    \centering
    \includegraphics[width=6.1in]{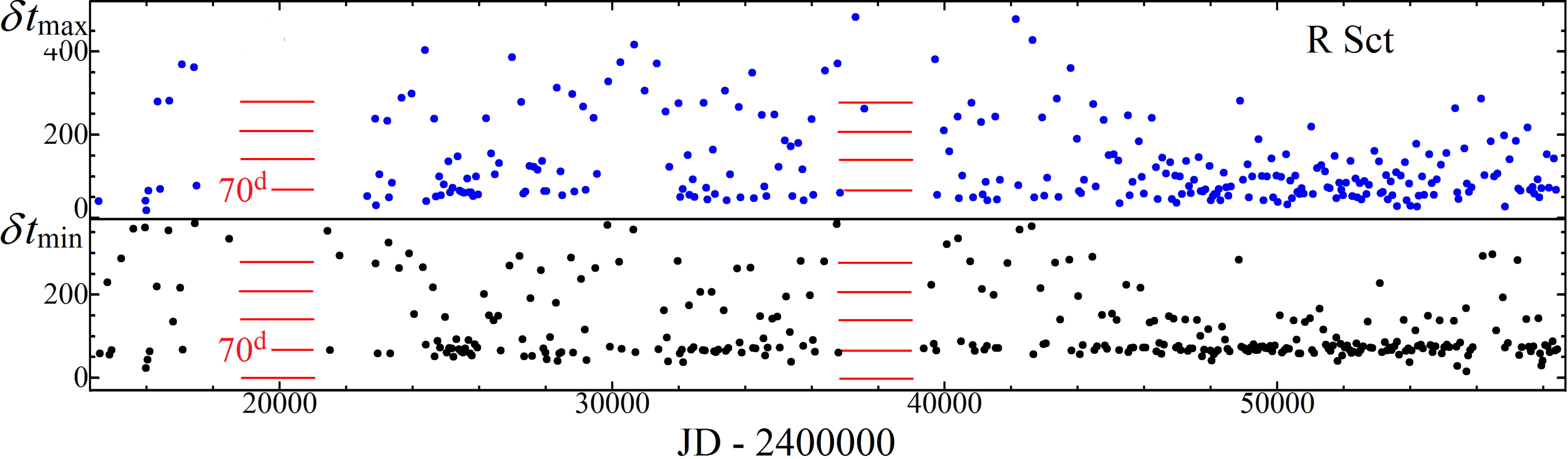}
    \caption[Fig.]{Dependence of the time intervals $\delta t$ for the maxima and minima. The horizontal red lines are multipliers of $70^{\rm d},$ an approximate period. For a constant period nad observations without gaps, all the points should be at a horizontal line corresponding to the basic period. The multiplied values correspond to absent (or very shallow) extrema.}
    \label{fig:6}
\end{figure}

The individual time intervals $\delta t$ were divided by a preliminary value of the period $P=71.74^{\rm d}.$ If these (non-integer) cycle counts $\delta E=\delta t/P$ were $\le 7,$ the values $\delta t$ were summed, as well as the rounded values $\delta J={\rm int}(\delta E+0.5).$ 
The ratio of these sums had given another estimate of the period $P_\delta=\sum(\delta t)/\sum(\delta J)=12401.65442
/175=70.713^{\rm d},$ close to the value from the periodogram. The values of $\delta t$ are shown in Fig. 6.

One may see apparent lines, what argues in a systematic period difference from that obtained. 
One may note a ''cell''-type structure of the phases.
We tried to change the trial period with a small step, to see the structure of the phase changes. Taking into account that the phase may cross the limiting values -0.5 and +0.5, we have shown the diagram in triple, formally increasing the phase from -0.5 to 1.5. Thus any minimum is shown in triple with a shift of 1. 

Finally, we adopted a value of $P=70.5^{\rm d}$
 It was used to compute the phases, even if the usually assumed period is twice larger. The initial epoch was arbitrary set to the moment of the first detected deep minimum:
\begin{equation}
Min.JD=2414814.02382+70.5\cdot E.
\end{equation}

In Fig. 7, the phases are shown according to the ephemeris (Eq. (3)). For comparison, we have shown ''artificaial'' moments of minima computed according to the table of interval-based ephemerids listed in the GCVS. While they are in a reasonable coincidence with our results, the ephemeris after JD\,2445000 is very different for the observed minima, possible, due to a miscomputation of the number of cycles.

The red line shows the period $P=70.74^{\rm d},$ which corresponds to the highest peak at the periodogram for $s=1,$ which is also seen for other $s.$
The decreasing black line corresponds to a half of the second significant peak of the periodogram of $138.35^{\rm d},$ also marked at Fig. 2. These two ''periods''  are not very good approximations because of phase shifts.

The strange situation with two ''periods'' argues for an absence of the true period. As the next approximation, one may suggest that one period $(\sim 70.8^{\rm d})$ may be switched to another $(\sim 69.8^{\rm d})$ and vice versa.

\begin{figure}
    \centering
    \includegraphics[width=6.1in]{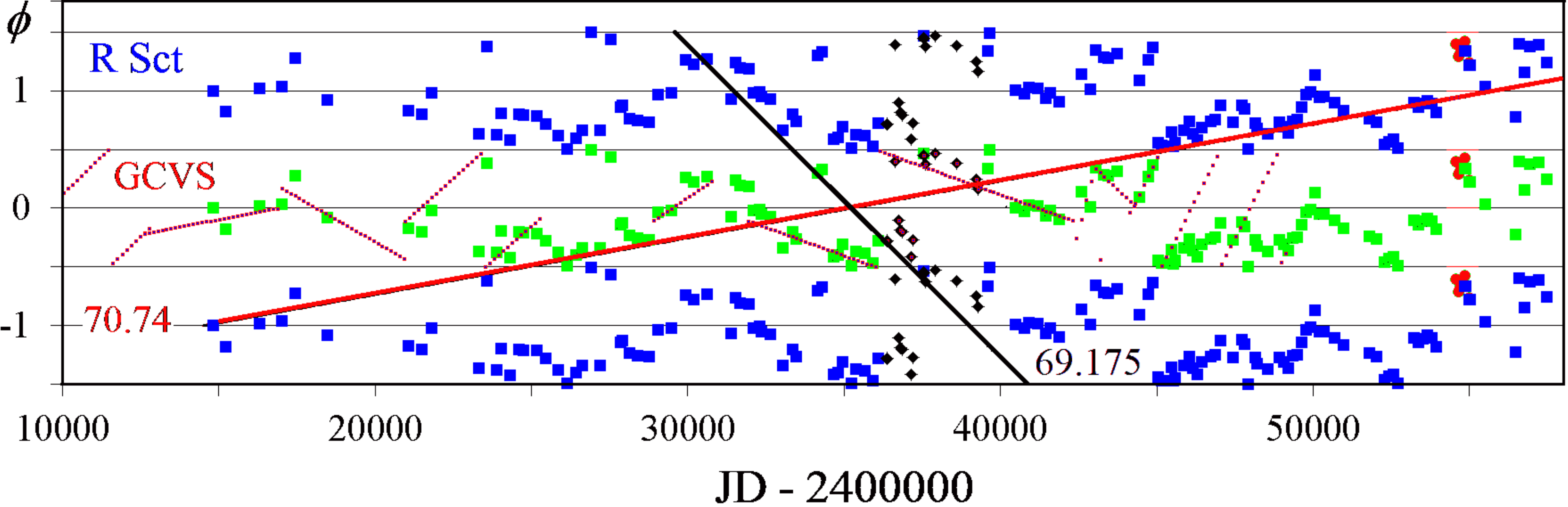}
    \caption{Dependence of phases of deep (green and blue points) and selected shallow (black) minima. 
    Red circles in the narrow interval JD\,2454604-55016 correspond to 7 data points compiled by H\"ubscher (2011) and do not contradict our findings.
    For suitability of looking for phase jumps, these data are shown in triple}
    \label{fig:7}
\end{figure}

\section*{Discussion}
As noted by some authors, R Sct is one of the most irregular stars in its class of pulsating variables. R Sct has such large variations in the depth of the minimums that it is often difficult to determine in the cycles of the formal period which minimums are primary (main) and which are secondary. Gillet (1992) writes that for R Sct each frequency peak shows a multi-component structure which is not encountered within power spectra deduced from hydrodynamic stellar models.This means that the pulsation of R Sct is never strictly periodic.

The multi-component structure of the frequency signal is characteristic of many stars. However, among this class of objects there are those that show practically "clean" peaks in a frequency ratio of 2: 1. According to the type of periodograms, we previously divided the stars into three groups: group I includes objects showing the periodogram typical form of RV Taurus stars, and the ratio of the periods of the two main peaks is indicated. Group II includes objects whose periodograms contain signs of multiperiodicity (Multi-p) or vice-versa, only one clear peak, instead of two (Single-p). Group III includes objects whose periodograms are highly noisy mainly due to the small number of observations. They do not show the typical details of RV-type stars. (Kudashkina L.S., 2020b; Kudashkina L.S., 2020a). According to this classification, R Sct can be placed in at least two different groups (1 and 2), depending on the interval of its light curve.

For the appearance of stochastic behavior with a small number of excited modes, two factors are required: resonant coupling of modes and nonlinear phase drift that interferes with mutual synchronization. Also, the linear coupling of modes with close frequencies plus the inertia of the medium can lead to stochastic behavior (Rabinovich, 1978). These conditions can be realized in R Sct, if, as indicated by a number of authors, this star is still on the AGB in the stage of a thermal-pulsation cycle of flare combustion of helium (Matsuura, M. et al., 2002). An additional inflow of energy from a flare violates the stationary self-oscillation regime, bringing the star into a state of low-dimensional deterministic chaos, that is, into a state on the verge of disappearance of regular self-oscillation regimes.

On the other hand, the violation of the regularity of the oscillations can be caused by the complex interaction of shock waves propagating in the photosphere and the extended atmosphere of the star. These effects can also be superimposed on the impact from the dust envelope (Gillet et al., 1989; Yudin et al., 2003). We previously studied the influence of shock waves on the light curve for long-period variables that are located on the AGB and, possibly, are closely related to RV Tauri - type  stars (Kudashkina \& Rudnitskij, 1994, 1988). It is also impossible not to take into account random changes in the period, for example, as a result of a change in the convection mode.

\section*{Conclusion}
Our work is based on R Sct observations published by the American Association of Variable Stars observers (AAVSO, Kafka 2020) for 120 years.

The periodogram analysis was carried out. The best period for the entire data set is $138.34^{\rm d}$ for $s= 1$. For $s=4$ elements for a phase light curve with a period $414.95^{\rm d}$ are obtained. Also, this period was refined to the value $415.03^{\rm d}$.

The scalegram analysis was carried out. Three peaks were revealed corresponding to the values of the periods 70, 135 and 284 days.

Thus, the main period is interrupted with phase shifts. Due to this, arises the apparent conclusion that the star has two periods and switches between them.

An alternate model for variability of R Sct is due to non-linear chaos (Buchler and Koll\'ath, 2003). It shows beat-like nearly periodic modulation of amplitudes and phases. Our analysis argues for ''switches'' between close periods with a characteristic time of $\sim 29$yrs.

\hspace{-0.7cm}{\bf Acknowledgements.} We acknowledge with thanks the variable star observations from the AAVSO International Database contributed by observers worldwide and used in this research. This work is in a frame of the international projects ''Inter-longitude astronomy'' (Andronov et al. 2003) and ''Astroinformatics'' (Vavilova et al, 2016).
\\

{\normalsize
\hspace{-0.7cm}{\bf References}
\begin{enumerate}[label={[\arabic*]},itemsep=-3pt]
\item Andronov I.L., 1994, Odessa Astron. Publ., 7, 49

\item Andronov I.L., 1997, Astronomy and Astrophysics. Suppl. Ser., 125, 207

\item  Andronov I.L., 2003, ASPC, 292, 391

\item  Andronov I.L., 2005, ASPC, 335, 37

\item  Andronov I.L., 2020, Knowledge Discovery in Big Data from Astronomy and Earth Observation, 1st Edition. Edited by Petr Skoda and Fathalrahman Adam. ISBN: 978-0-128-19154-5. Elsevier, 2020, p.191-224

\item  Andronov I.L.,
et al., 2003, Astronomical \& Astrophysical Transactions, 22, 793

\item  Andronov I. L., Chinarova L. L., 2003, Astronomical Society of the Pacific Conf. Ser. 292., 401 
\item Andronov I.L., Baklanov A.V., 2004,  Astronomical School's Report, 5, 264-272

\item Andronov I.L., Baklanov A.V., 2004, https://soft.softodrom.ru/ap/Multi-Column-View-MCV-p7464

\item Andronov I.L., Breus V.V., Kudashkina L.S., 2020,
Mathematical Modelling of Astrophysical Objects and Processes, in: "Development Of Scientific Schools Of Odessa National Maritime University": Collective monograph. Riga, Baltija Publishing, pp.3-29

\item Andrych K.D., Andronov I.L., Chinarova L.L., Marsakova V.I.m 2015, Odessa Astron. Publ., 28, 158 

\item  Andrych K.D., Andronov I.L., Chinarova L.L., 2020a, Journal of Physical Studies, 24, 1902

\item  Andrych K.D., Tvardovskyi D.E., Chinarova L.L., Andronov I.L., 2020b, Contributions of the Astronomical Observatory Skalnaté Pleso, 50, 557

\item Buchler J.R., Koll\'ath Z., 2003< in: Y.  Nakada et al.  (eds.), Mass-Losing Pulsating Stars and their Circumstellar Matter,  Kluwer Academic Publishers, p.  59-66 

\item Chinarova L.L., Andronov I.L., 2000, Odessa Astron. Publ., 13, 116

\item  Espin T.E., 1890, MNRAS, 51, 11, 1890MNRAS..51...11E

\item  Foster G., 1996, Astronomical Journal, 112, 1709

\item  Gillet, D. et al., 1989, Astronomy and Astrophysics 215, 316

\item  Gillet, D., 1992, Astronomy and Astrophysics 259, 215

\item H\"ubscher J., 2011, Open European Journal on Variable Stars, 131, 1

\item  Jura, M., 1986, ApJ 309, 732

\item  Kafka, S., 2020, Observations from the AAVSO International Database,\\ https://www.aavso.org

\item  Kudashkina L.S., et al., 1998, Proc. 29th Conf. Var. Star. Res., Brno, Czech Rep. 126 1998vsr..conf..126K

\item  Kudashkina L.S., 2019, Astrophysics, 62, 4, 556.

\item  Kudashkina L.S., 2020a, Odessa Astron. Publ., 33, 34.

\item  Kudashkina L.S., 2020b, Annales Astronomiae Novae, 1, 199; 2020AANv....1..199K 

\item  Kudashkina L.S., 2020c, Astronomical and Astrophysical Transactions, 31, 4, 451

\item  Kudashkina L.S., Rudnitskij G.M., 1988, Peremennye Zvezdy, 22, 925

\item  Kudashkina L.S., Rudnitskij G.M., 1994, Odessa Astron. Publ., 7(1), 66

\item  Kudashkina L. S., Andronov I.L., 1996, Odessa Astron. Publ., 9, 108

\item  Kudashkina L.S., Andronov I.L., 2017a, Odessa Astron. Publ., 30, 93.

\item  Kudashkina L.S., Andronov I. L., 2017b, Częstochowski Kalendarz Astronomiczny, ed. Bogdan Wszołek, 14, 283; ADS: 2017CKA....14..283K . arXiv:1711.09029

\item  Kudashkina L.S., Andronov  I.L., Grenishena L.V., 2013, Częstochowski Kalendarz Astronomiczny, ed. Bogdan Wszołek, 9, 211

\item  Mantegazza, L., 1991, Astronomy and Astrophysics. Suppl. Ser., 88, 255

\item Marsakova V.I., Andronov I.L., 1996, Odessa Astron. Publ., 9, 127 

\item  Matsuura M., et al., 2002, Astronomy and Astrophysics 387, 1022

\item  Rabinovich M.I., 1978, UFN, 125, 1, 123

\item  Samus N.N., et al., 2017,
Astronomy Reports, 61, 80; 
http://sai.msu.su/gcvs/gcvs/

\item Vavilova I.B., Yatskiv Y.S., Pakuliak  L.K., Andronov  I.L., et al., 2016, Proceedings of the International Astronomical Union 12 (S325), 361

\item  Yudin R.V. et al., 2003, Astronomy and Astrophysics 412, 405
\end{enumerate}}
\end{document}